\documentclass{appolb}%
\usepackage{graphicx}
\usepackage{amsmath}

\def\vec#1{\ensuremath{\mathchoice
                     {\mbox{\boldmath$\displaystyle#1$}}
                     {\mbox{\boldmath$\textstyle#1$}}
                     {\mbox{\boldmath$\scriptstyle#1$}}
                     {\mbox{\boldmath$\scriptscriptstyle#1$}}}}

\begin{document}
%
%
\title{The effective QCD running coupling constant and a 
Dirac model for the charmonium spectrum
}
\author{M. De Sanctis \footnote{mdesanctis@unal.edu.co}
\address{Universidad Nacional de Colombia, Bogot\'a, Colombia }
\\
}
\maketitle
\begin{abstract}
The  QCD \textit{effective charge} extracted from the experimental data is
 used to construct
the vector interaction of a Dirac relativistic model for the charmonium spectrum.
The process required to fit the spectrum is  discussed and 
the relationship with a previous study of 
the vector interaction is analyzed.

\end{abstract}
\section{Introduction}\label{intro}
In a series of previous works the author developed a Dirac relativistic
quark-antiquark  model
to study the spectrum of charmonium and, possibly, of other mesons.
In particular, in Ref.   \cite{localred} the relativistic   
reduced Dirac-like equation (RDLE) of the model was introduced.
This equation is written  in the coordinate space in a local form.
%
An accurate calculation of the charmonium spectrum was performed using a small
 number of free parameters in Ref. \cite{rednumb}.
Furthermore,
in a subsequent work \cite{relvar}, the Lorentz structure of the interaction terms was
studied in more detail, 
developing a covariant form of the  same RDLE.

In  this model, 
a specific  form of the  \textit{regularized vector interaction} has been used.
That interaction had been
 introduced and studied previously in Ref. \cite{chromomds}.
We highlight here that a vector interaction alone is not sufficient to give 
an accurate reproduction of the charmonium spectrum.
To this aim, the contribution of a
\textit{scalar interaction}  has been always included in the interaction of the RDLE.
In this respect,  
the scalar  interaction was studied in more detail 
in another work \cite{scalint},
also considering the possibility of using
a \textit{mass interaction}.
In the same work the scalar and mass interactions have been tentatively
related to the excitation of the first scalar resonances of the hadronic spectrum.
In the following, we shall denote the content of all these works
(and the corresponding results)
as \textit{previous calculations}, performed with the RDLE.

\vskip 0.1 truecm
In the present work we go back to the study of the vector term of the interaction
exploring a possible relationship between this interaction term  and
the Quantum Chromo-Dynamics 
(QCD) effective \textit{running} strong coupling constant $\alpha_S(Q)$,
where $Q$ represents, as usual,  the quark vertex momentum transfer. 
In particular we shall consider, for $\alpha_S(Q)$,
the \textit{effective charge}
 $\alpha_{g1}(Q)$  that was extracted from the experimental data 
using the generalized Bjorken sum rule.
The procedure of extraction and the theoretical analysis have been performed 
in different works \cite{Deur05, Deur08, Deur22}
to which we refer specifically for the present study.
Furthermore,
in the extensive review on  the QCD running coupling constant 
$\alpha_S(Q)$ 
\cite{Deurrev23}, in the previous review \cite{Deurrev16} and
in the references quoted in these papers,
the theoretical and phenomenological properties of $\alpha_{g1}(Q)$ 
are also analyzed in detail.

\vskip 0.1 truecm
As shown in the previously cited works
\cite{Deur05,Deur08,Deur22,Deurrev23,Deurrev16},
the extracted  $\alpha_{g1}(Q)$ coincides, at high momentum transfer 
with the predictions of perturbative QCD for $\alpha_S(Q)$. 
At low momentum transfer, 
$\alpha_{g1}(Q)$ can provide
a reliable definition  of the strong coupling constant,  
offering a potentially relevant tool for the study of the
\textit{nonperturbative} hadronic phenomena,
such as the emergence of hadronic mass, quark confinement
and hadron spectroscopy.
In this respect, a crucially relevant property of $\alpha_{g1}(Q)$
is that this quantity does not present any low $Q$ divergence but 
``freezes'' as $Q\rightarrow 0$.
In other words, in this limit,  it loses
its $Q-$dependence.

Some care must be exercized considering that different forms of 
\textit{effective charges} can be introduced in relation to
different observable hadronic quantities.
In the present work we take specifically $\alpha_{g1}(Q)$ due to the great number
of high precision experimentally extracted data
that allow to
construct, without numerical
uncertainties, a suitable vector interaction  for  our RDLE.

In Refs. \cite{Deurrev23,Deurrev16} the authors also discuss
the form of $\alpha_S(Q)$ in some nonperturbative
approaches to QCD.
Due to the interest
for the development of the present work, 
we recall that
the Holografic Light-Front QCD  gives a 
``freezing'' coupling constant \cite{Deurrev23}   
$\alpha_{HLF}(Q)$.
Also, in the Richardson
model \cite{richar79},  a static potential for the 
constituent-quark interaction is introduced. 
This potential  grows linearly with
the quark distance.
From this potential one can formally obtain an effective coupling constant
$\alpha_{Rich}(Q)$
that, however, is 
\textit{divergent} as $Q\rightarrow 0$.
Furthermore,
in Refs. \cite{Deur05,Deur08}
a comparison of $\alpha_{g1}(Q)$
with the coupling constant of the 
Godfrey-Isgur  constituent-quark model \cite{godf_isg85} is given.

Finally, we highlight here that, 
as explained in the detailed analysis of Refs. \cite{Deurrev23,Deurrev16},
 the connection between an expression of
$\alpha_S(Q)$  ``in accordance'' with QCD and the quark interaction 
for the hadronic bound states, is not univocally defined and
still represents a challenge for theoretical physics.


\vskip 0.2 truecm
Taking into account the complexity of the problem,
in the present work we shall revise the previously developed vector interaction
of Refs. \cite{localred,rednumb,relvar,chromomds,scalint},
deriving from the quantities introduced there 
the (possibly) related form of $\alpha_S(Q)$.
Then we shall use the effective coupling $\alpha_{g1}(Q)$
of Refs. \cite{Deur05,Deur08,Deur22} 
to construct, with some modifications, the vector potential 
for the model.

Finally,
we point out that, by means of our RDLE,  a truly \textit{relativistic} model
is constructed. 
In this model the vector interaction
and the scalar (or mass) interaction can be treated \textit{separately},
allowing for a \textit{separate} study of their structure.
In particular, in the present work, we shall focus our attention 
on the vector interaction.\\
We recall that, on the contrary,
in the nonrelativistic studies, the two interactions
give rise 
(at least at the leading order in the nonrelativistic expansion) 
to a \textit{unique} potential,
in which the two contributions cannot be easily disentangled.

\vskip 0.2 truecm
The remaider of the paper is organized as follows.
In the next Subsect. \ref{symbnot} the notation and conventions used in the work
are introduced.
In Sect. \ref{vectmcspace}, we study the theoretical connection between
the running coupling constant, as a function of the momentum transfer $Q$, 
with the vector interaction potential.
In Sect.  \ref{gqrm}, we analyze from the (new) point of view of this paper 
our previuos calculations preformed with the RDLE.
In Sect.  \ref{alphag1}, we develop the construction of the interaction vector potential 
by using the experimentally extracted $\alpha_{g1}(Q)$.
Finally, in Sect. \ref{charmspectr}, the charmonium spectrum is calculated and displayed.
The role of the different parameters is analyzed and some general considerations 
about the whole problem are given.

\vskip 2.5 truecm
\noindent


\subsection{Notation and conventions}\label{symbnot}

The following notation and conventions are used in the paper.

\begin{itemize}
\item  
The invariant product between four vectors is standardly written as: 
$V^\mu U_\mu= V^\mu U^\nu g_{\mu \nu}=V^0 U^0- \vec V \cdot \vec U$.
\item
The lower index $i =1,2$ represents the \textit{particle index},
referred to the quark ($q$) and to the antiquark ($\bar q$).
\item
We shall use, for each quark,  the four Dirac matrices $\gamma_i^\mu$.
\item
The vertex 4-momentum transfer will be denoted as $ q^\mu=(q^0, \vec q)$.
\item
We shall
neglect  the retardation contributions, setting $q^0=0$ for the time component
of the 4-momentum transfer.
This approximation is consistent with the use of the Center of Mass Reference Frame
for the study of the $q \bar q$ bound systems.
\item
In consequence, the \textit{positive} squared four momentum transfer $Q^2$
takes the form $Q^2=-q_\mu q^\mu= \vec q^2$, that is $Q=|\vec q|$.
%
\item
The quantities $\alpha(Q)$, $G(Q)$, $V^V(r)$ and $U^V(r)$
that will be introduced in the paper,  are used, \textit{with no label},
 in general expressions.
\item
To indicate the model 
to which these quantities are referred, a specific label is added:
$Coul$ for the pure Coulombic case, $pr$ for
the previous calculations with the RDLE and
$g1$ for the effective charge extracted from the experimental data.
The quantity $\alpha_V(0)$ will be also introduced
in Sect. \ref{charmspectr}.
\item
The subindex $X$ will be used to denote, 
for the parameters $\bar V_X$ and $r_X$,
the scalar ($X=S$) or mass ($X=M$) character of the corresponding interaction.
\item
Finally, throughout the work, we use the standard natural units, 
that is $\hbar=c=1$.\\
\end{itemize}

\vskip 1.0 truecm
\section{ The vector interaction in momentum and coordinate space}\label{vectmcspace}
Our RDLE \cite{localred,rednumb} 
has been formulated in the coordinate space.
In order to introduce into this model the momentum dependent 
running coupling constant $\alpha_S(Q)$, 
it is strictly necessary to establish the connection 
between the coordinate space and the
momentum space interaction.
We write, \textit{in general}, the momentum dependence of the vector strong interaction 
(apart from the standard $1/Q^2$ factor) in the form
\begin{equation}\label{momdep}
\alpha(Q)=\alpha (0) G(Q)
\end{equation}
where $ \alpha (0)$ is a \textit{truly} constant, adimensional quantity 
that  ``represents the strength'' of the vector interaction. 
Furthermore,
$G(Q)$ is a decreasing, positive, 
function of the momentum transfer  $Q$ that satisfies the condition
$G(0)=1$.
The momentum dependence of $\alpha(Q)$ can be related, at a fundamental level,
to the running of the QCD coupling constant,
identifying $\alpha(Q)$ with the strong coupling constant $\alpha_S(Q)$.
In phenomenological quark models, 
as, for example, in our previous calculations,
we can say that the function $G(Q)$ takes phenomenologically into account 
the structure of 
the interacting, nonpoint-like, quarks.
Its physical meaning, within different models,
will be   analyzed  in more detail in the following of the paper.


\vskip 0.2 truecm
By means of Eq. (\ref{momdep}),
the tree-level vector interaction in the momentum space,
for a $q \bar q$ system,
 can be written, in general, as
\begin{equation}\label{intmom}
{\cal W}^V(Q)= -{\frac 4 3} {\frac {4 \pi} {Q^2} } \alpha(0)  G(Q)
\gamma_{1}^\mu \gamma_2^\nu g_{\mu \nu}
\end{equation}
where
$4/3$ represents   the color factor
in the $q \bar q$ case; 
$ \alpha (0)$ and $G(Q)$ have been
introduced in Eq. (\ref{momdep}). 
Performing the Fourier transform one obtains the corresponding expression 
in the coordinate space
\begin{equation}\label{fourtrans}
{\cal W}^V(r)=\int {\frac {d^3 q} {(2\pi)^3}}
 \exp{(i \vec q \cdot\vec r)} {\cal W}^V(Q)
\end{equation}
Multiplying the previous expression by $\gamma^0_1 \gamma ^0_2$
from the left, one obtains,
the two-body vector interaction 
$W^V_{(2)}$ introduced in Eq. (10) of Ref. \cite{rednumb}
for the calculations in the Hamiltonian Dirac form.

In particular, the two-body interaction potential in the coordinate space
is given by the following Fourier transform
\begin{equation}\label{potfourtrans}
V^V(r)=  -{\frac 4 3} 
 \int {\frac {d^3 q} {(2\pi)^3}}
 \exp{(i \vec q \cdot\vec r)}
{\frac {4 \pi} {Q^2} }  \alpha (0) G(Q)~,\\
\end{equation}
where $V^V(r)$ is the vector (two-body) interaction potential, 
denoted as $V^{int}(r)$ in Eqs. (12) and (14)  of Ref. \cite{rednumb}.
%
In the first place, we recall that,
in the case of a constant $G(Q)$,
one goes back to a standard Coulombic interaction.
More precisely, for 
 $G_{Coul}(Q)=1$, one would obtain in the coordinate space
the pure Coulombic potential
\begin{equation}\label{purecoul}
V_{Coul}^{V }(r)=  -{\frac 4 3} {\frac  { \alpha_{Coul} (0)}  {r}}~.
\end{equation}
This potential is not able to reproduce with good accuracy the charmonium spectrum.
Furthermore, the choice $G(Q)=G_{Coul}(Q)=1$ is not in agreement with the
QCD phenomenology, being completely ignored the running of the coupling constant.

In the following Sect. \ref{gqrm} we shall discuss $G_{pr}(Q)$, corresponding
to the potential
$V_{pr}^{V }(r)$  that was introduced in our previous works \cite{rednumb,scalint}.
In Sect. \ref{alphag1} we shall study the case of $\alpha_{g1}(Q)$ extracted from 
the experimental data.
In any  case, the interaction potential in the coordinate space
 is obtained by means 
of the Fourier transform of Eq. (\ref{potfourtrans}).

\section{The quantity $G_{pr}(Q)$ of our previous calculations }\label{gqrm}
The impossibility of reproducing accurately the charmonium spectrum 
with a pure Coulombic potential required to use, 
in Ref. \cite{rednumb}, a model of the vector interaction that was previously 
introduced in Ref. \cite{chromomds}.
In this model the quarks are considered as \textit{extended} sources of 
the chromo-electric field.\\
After many trials with different analytic functions,
an accurate reproduction of the charmonium spectrum 
has been obtained with a Gaussian color charge distribution for each quark:
\begin{equation}\label{rhogauss}
\rho( x)={\frac {1} {(2 \pi d^2)^{3/2} }}
\exp\left(-{\frac {\vec x^2} {2d^2} } \right)~.
\end{equation}
This distribution gives, in the momentum space, the following vertex form factor
\begin{equation}\label{ffgauss}
F(Q)=\exp (-{\frac {Q^2 d^2} {2}})
\end{equation}
Considering  one form factor for each quark vertex, one obtains
for the function $G(Q)$  introduced in  Eq. (\ref{intmom}),
the following expression, specific of our previous calculations:
\begin{equation}\label{gqgauss}
G_{pr}(Q)=[F(Q)]^2= \exp (- Q^2 d^2 )~.
\end{equation}
For this model, developed in our previous calculations, 
we have the (true) constant $\alpha_{pr}(0)=\alpha_V$ 
that was introduced in Refs.  \cite{rednumb,scalint}.
As anticipated at the beginning of the previous section,
we can say that, \textit{within this model},
 the quantity $\alpha_{pr}(Q)=\alpha_{pr}(0) G_{pr}(Q) $
%
\textit{defines} an effective strong running
coupling constant $\alpha_S(Q)$.
Furthermore, we observe that $\alpha_{pr}(Q) $,
with $G_{pr}(Q)$ of Eq. (\ref{gqgauss}),   
is a function without singularities that ``freezes''
(\textit{i.e.} goes to a constant limit)
as $Q\rightarrow 0$.

By performing the Fourier transform defined in Eq. (\ref{potfourtrans}),
with $G_{pr}(Q)$ of Eq. (\ref{gqgauss}),
one obtains the interaction potential  in the following analytic form
\begin{equation}\label{vintgauss}
 V_{pr}^V( r)=- {\frac 4 3}  {\frac {\alpha_{pr}(0)} {r}}
\text{erf} \left(  {\frac {r} {2d}  }\right)~.
\end{equation}
In Eq. (17) of Ref. \cite{rednumb}
the same result,
denoted there as $V^{int}(r)$,
was obtained by means of a different procedure
completely developed in the coordinate space.
Note that the potential of Eq. (\ref{vintgauss}) is
\textit{regular} for $r \rightarrow 0$.
More precisely, we have
\begin{equation}\label{vpr0}
V_{pr}^V( 0)=- {\frac 4 3}  {\frac {\alpha_{pr}(0)} {d}}
{\frac {1} {\sqrt{\pi}}}~.
\end{equation}
This result was given in Eqs. (13) and (16) of Ref. \cite{rednumb}.

We recall that  also a positive constant term,
denoted as $\bar V_V$,
 is frequently introduced in
quark models to improve the reproduction of the experimental spectra. 
In our previous calculations,
as shown in Eq. (13) of Ref. \cite{rednumb},
  we fixed this constant in the following way:
\begin{equation}\label{vbarvdef}
\bar V_V=-V_{pr}^V( 0)~.
\end{equation}
With this assumption, the constant $\bar V_V$
represents the positive zero-point quark self-energy that, added to the
interaction term of Eq. (\ref{vintgauss}), 
gives a total vector potential that is vanishing at $r=0$ and
approaches the maximum value  $\bar V_V$ as $r\rightarrow \infty$.

\vskip 0.2 truecm

As discussed above, the parameters of the vector interaction,
in our previous calculations, are $d$ and $\alpha_{pr}(0)$.
Their   numerical values  
 were obtained by fitting 
the resonance masses of the charmonium spectrum.
The following numerical values 
were obtained:
$ d=(0.1526) ~0.1511~ fm $
corresponding to 
$\lambda=1/d= (1.293) ~ 1.306~ GeV$ 
and 
$\alpha_{pr}(0)= (1.864) ~1.838$
where the first values (in brackets) are those of Table II of Ref. \cite{rednumb}
and the second ones are those of Table II of  Ref. \cite{scalint}.
In the latter case an updated set of charmonium resonance masses
\cite{pdg22} were used
to determine the values of $d$ and $\alpha_{pr}(0)$.
In the remainder of this work, we shall consider only the second group of values.

\vskip 0.2 truecm

Incidentally,
these results can be compared with HLF QCD that gives,
for the effective running coupling constant \textit{exactly}
the same analytic expression:
\begin{equation}\label{alphahlf}
\alpha_{HLF}(Q)= \alpha_{HLF}(0)  \exp[- {\frac {Q^2} {(2 \kappa)^2}}]~.
\end{equation}
The numerical value is $2 \kappa=1.046\pm 0.048~ GeV$, as given in Ref. \cite{Deurrev23}.
This value has the same order of magnitude as $\lambda$ of our model.

\vskip 0.2 truecm
\noindent

\section{The use of $\alpha_{g1}(Q)$}\label{alphag1}
In this section we analyze the possibility of using
the quantity $\alpha_{g1}(Q)$, extracted from the experimental data,
to construct the vector interaction potential.
In the first place, considering the results of Refs
\cite{Deur05,Deur08,Deur22},
we write
\begin{equation}\label{alphag1first}
\alpha_{g1}(Q)= \alpha_{g1}(0) G_{g1}(Q)
\end{equation}
where one would have $ \alpha_{g1}(0) =\pi$ 
(this numerical value will be discussed in the following).
Then,
in order to perform (numerically) the Fourier transform 
of  Eq. (\ref{potfourtrans}),
required for the calculation of the vector potential, 
we parametrize $G_{g1}(Q)$
with a continous analytic function,
 in the following way:
\begin{equation}\label{gg1param}
G_{g1}(Q)= aA(Q)+(1-a)B(Q)
\end{equation}
where the two momentum dependent functions $A(Q)$ and $B(Q)$
satisfy the condition
\begin{equation}\label{cond}
A(0)=1, ~~B(0)=1~.
\end{equation}
In more detail, we take these functions in the form:
\begin{equation}\label{aq}
A(Q)= \exp(-Q^2 d_a^2)
\end{equation}
and
\begin{equation}\label{bq}
B(Q) ={\frac {1+c_0 \alpha_b \ln (x_b)}  
            {1+c_0 \alpha_b \ln (x_b+ Q^2 (\eta(Q))^2)} }
\end{equation}						
with
\begin{equation}\label{bqadd}
\begin{aligned}
&\eta(Q)=\eta_0+ b Q~,\\
&c_0=(11-n_f{\frac 2 3}){\frac {1} {4\pi}}~.\\ 
\end{aligned}
\end{equation}

\vskip 0.5 truecm
\noindent
The total function of Eq. (\ref{gg1param})
has been fitted to the experimentally extracted data 
\cite{Deur05,Deur08,Deur22},
from $Q=0$ to $Q=50~ GeV$,
obtaining the following values for the parameters of Eqs. 
(\ref{gg1param} - \ref{bqadd}):
$a=0.35415$, 
$d_a=0.1611~ fm$, 
$\alpha_b=1.395$, 
$x_b=0.9164$,
$\eta_0=0.7385 ~GeV^{-1}$, 
$b=1.479~GeV^{-2}$ and
$n_f=6$.
%
In the parametrization displayed above
$B(Q)$ of Eq. (\ref{bq}) is related to the low momentum behavior
of $\alpha_{g1}(Q)$, while $B(Q)$ takes into account the high momentum 
logarithmic terms, peculiar of perturbative  QCD. 
However, we point out that our parametrization does not pretend to have
a specific physical meaning but has been introduced 
essentially to perform the numerical calculation.

\begin{figure}
{%
  \includegraphics{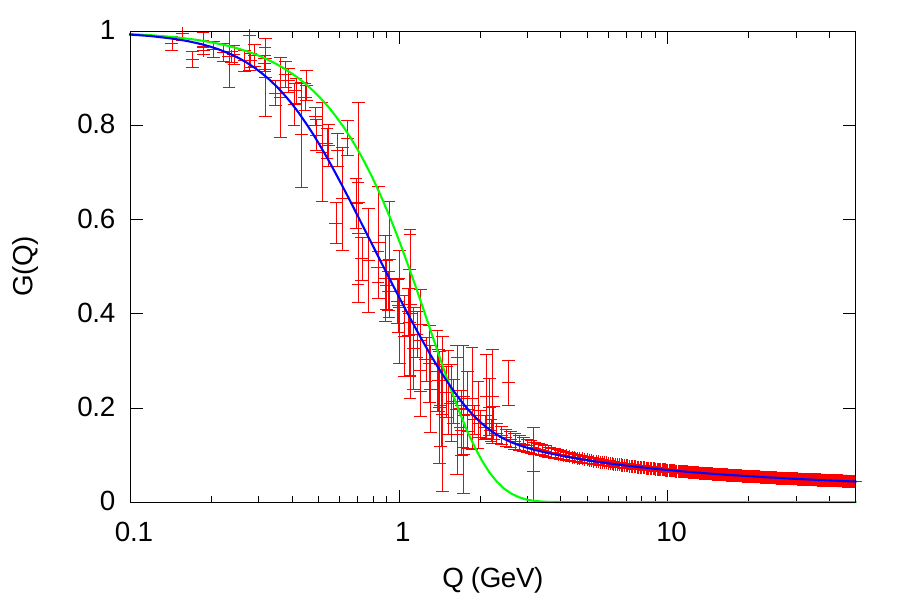}
}
\vspace{0.01cm}
\caption{
The function $G(Q)$ introduced in Eqs. (\ref{momdep}) and (\ref{intmom}). 
The points with error bars, in red, represent the experimentally extracted
$g1$ data; 
the blue continous line represents $G_{g1}(Q)$, that is  
 the fit of Eq. (\ref{gg1param}) to these data.
The green continous line represents $G_{pr}(Q)$ of our previous calculations,
given by Eq. (\ref{gqgauss}).
}
\label{fig_1}       
\end{figure}

The experimentally extracted data, the corresponding fit for $G_{g1}(Q)$
and $G_{pr}(Q)$ of   Eq. (\ref{gqgauss})  are shown in Fig. \ref{fig_1}.
In this figure, the  sources of the experimentally extracted data are not differentiated.
For more details regarding this point, the reader is referred to the works 
\cite{Deur05,Deur08,Deur22,Deurrev23}.

\vskip 0.2truecm

The coordinate space potentials are obtained by means of Eq. (\ref{potfourtrans}).
In particular, for the experimentally extracted data, we use the 
parametrization of $G_{g1}(Q)$ given in Eq. (\ref{gg1param})
with the functions $A(Q)$ and $B(Q)$ defined in Eqs. (\ref{aq}) and (\ref{bq}),
respectively.
The calculation is performed analytically for $A(Q)$
and numerically for $B(Q)$.

\vskip 5.0 truecm

\begin{figure}
{%
  \includegraphics{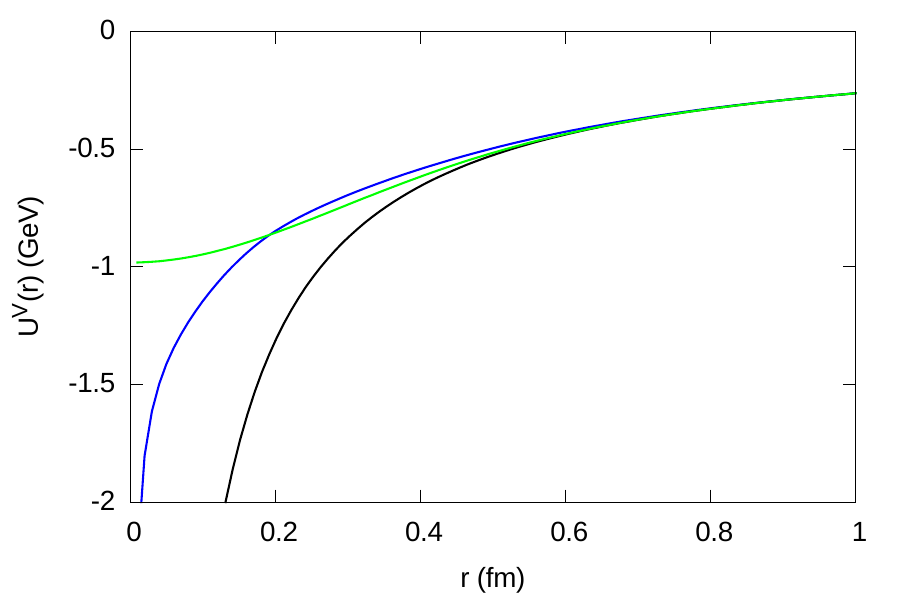}
}
\vspace{0.01cm}
\caption{
The coordinate space function $U^V(r)$ of Eq. (\ref{udef}). 
The blue line,  $U_{g1}^V(r)$,   is obtained from the fit of the experimental data;
the green line, $U_{pr}^V(r)$   is given by the potential of the previous model;
the black line, $U_{Coul}^V(r)$ represents the pure Coulombic case.
}
\label{fig_2}       
\end{figure}

In order to display graphically the coordinate space potentials,
we divide the potentials by $\alpha(0)$, introducing
the following coordinate space function
\begin{equation}\label{udef}
U^V(r)={\frac {V^V(r)} {\alpha (0)} }~.
\end{equation}
This function is plotted in  Fig.  \ref{fig_2}.
In more detail, in this figure,  we display:
\begin{itemize}
\item $U_{g1}^V(r)$,
obtained from the fit of the experimentally extracted  data;
\item  
$U_{pr}^V(r)$ given by Eq. (\ref{vintgauss});
\item
$U_{Coul}^V(r)$ that is given by the pure Coulombic potential
of  Eq. (\ref{purecoul}).
\end{itemize} 
We note that, as $r \rightarrow \infty$, the three functions 
have the same Coulombic behavior.
As $ r \rightarrow 0$, 
$U_{pr}^V(r)$ takes the finite value determined by Eq. (\ref{vpr0});
numerically, this value is
$U_{pr}^V(0)=-0.9824~GeV$.
This regularization of the potential is given by 
the fastly decreasing function $G_{pr}(Q)$.
On the other hand, $U_{g1}^V(r)$ diverges as $r\rightarrow 0$, 
with a slower rate than $U_{Coul}^V(r)$.
In this respect, we observe that the function
 $B(Q)$ of $G_{g1}(Q)$ does not  decrease sufficiently fast, as $Q\rightarrow \infty$,
to regularize the corresponding coordinate space potential when $r\rightarrow 0$.

\vskip 0.5 truecm
\noindent

\section{The charmonium spectrum}\label{charmspectr}
We can now try to reproduce the charmonium spectrum 
with the vector potential given by $U_{g1}^V(r)$.

\vskip 0.2 truecm
\noindent
The technique for solving the RDLE
and the fit procedure
are exactly the same as in Refs. \cite{rednumb,scalint}.
For the charmonium spectrum we use here the  experimental data \cite{pdg22}.
\vskip 0.2 truecm
For the quality of the fit, as in \cite{scalint}, 
we define
\begin{equation}\label{qual}
\Theta=\sqrt{ {\frac {\sum_k(E_k^{th} -M_k^{exp} )^2} {N_d}} }~,
\end{equation}
where $E_k^{th}$ and $M_k^{exp}$ respectively represent
the result of the theoretical calculation and the experimental value 
of the mass,
for the $k$-th resonance and $N_d=16$ is the number of the fitted resonances.

\vskip 0.2 truecm
We point out that the model, to reproduce accurately the spectrum,
necessarily includes also
 a scalar $(S)$
\cite{localred,rednumb},
or mass $(M)$ \cite{scalint}, interaction.

We  have started the analysis trying to \textit{fix} the vector interaction strength
at the value $\alpha_{g1}(0)=\pi$, 
as given in Refs. \cite{Deur05,Deur08,Deur22}. 
But this choice
did not allow to reproduce 
accurately the charmonium spectrum. In this respect, 
many trials have been performed modifying the form of the scalar or
mass potentials. 
We have also tried to modify the form of $G(Q)$ but,
in any case, the fit of the charmonium spectrum refused the value
$\alpha_{g1}(0)=\pi$ 
of the vector interaction strength. 

Subsequently, this quantity, that we denote from now on 
as $\alpha_V(0)$, has been left as a
\textit{free parameter of the fit}.
This choice has allowed an acceptable reproduction of the charmonium spectrum,
as shown in Table \ref{tabspectr},
where the theoretical and the experimental values 
of the resonance masses are displayed.
The values of the parameters used for the interaction
are given in Table \ref{tabpar}.

In particular,
for the mass of the quark we have taken the same value of the previous works
\cite{rednumb},\cite{scalint}, that is $m_q=1.27~GeV$.
This value represents the ``running" charm quark mass 
in the $\overline{MS}$ scheme \cite{pdg22}.

As discussed before, $\alpha_V(0)$ is determined by the fit to the spectrum.
Comparing the  results obtained for $\alpha_V(0)$  with $\alpha_{g1}(0)=\pi$,
we have: $\alpha_V(0)=0.65~ \alpha_{g1}(0)$ and $\alpha_V(0)=0.62 ~\alpha_{g1}(0)$,
when the scalar or mass interaction are used, respectively.
As discussed in the introduction, the nonunivocal definition of the effective charge,
that affects particularly the low $Q$ region, can explain why 
the value $\alpha_{g1}(0)=\pi$
is not adequate for obtaining a suitable bound state quark interaction
for our calculation.

With respect to   Ref. \cite{scalint}, here the additional constant 
of the vector interaction $\bar V_V$
is considered as a completely free parameter:
the vector interaction obtained from $G_{g1}(Q)$ does not allow 
to relate $\bar V_V$
to the quark vector self-energy.

Following the phenomenological model  discussed in \cite{scalint}, 
we have \textit{fixed} the constant
$\bar V_X$, for both the scalar ($X=S$) and the mass ($X=M$) interaction
at the value $\bar V_X=0.7350~GeV$.
Also for the distance parameters $r_X$, the same values of \cite{scalint}
have been used,
as shown in Table \ref{tabpar}. 

Analyzing in more detail the obtained results for the spectrum, 
we note that the quality of the fit is slightly worse here than in Ref. \cite{scalint}.
For the parameter $\Theta$ defined in  Eq. (\ref{qual}),
we have  here $\Theta= 36.0 ~MeV$ and $\Theta=38.0~MeV$ for the Scalar and Mass
interaction, respectively. 
In Ref. \cite{scalint}, the corresponding values were $\Theta= 13.4~MeV$ and $\Theta=12.8~MeV $.
 The quality of the fit can be improved
if the parameters $V_X$ and $r_X$ are left as free parameters.
We decided to fix these parameters at the same values of  Ref. \cite{scalint}
to show that the vector potential obtained from $G_{g1}(Q)$ is compatible
with the model for the scalar and mass interactions studied in  Ref. \cite{scalint}
without changing their parameters.

For completeness,
we also note that, as in \cite{scalint}, the model is unable to reproduce the resonance 
 $\chi_{c0}(3915)$. 
The new experimental data \cite{pdg22} give, for this resonance, a mass of
$3921.7 \pm 1.8 ~MeV$. 
Our model, taking the quantum numbers $2^3P_0$,   
gives  the  mass  values of $3857~MeV$ and $3846~MeV$, for the $S$ 
and the $M$ interactions, respectively.  
Our model and other quark models give a wrong order for the masses of 
this resonance and its partner $\chi_{c1}(3872)$.

\vskip 0.2 truecm
We conclude this paper with the following considerations.
The momentum dependence of the QCD experimentally extracted $\alpha_{g1}(Q)$
gives a vector interaction potential that is compatible with our quark model
based on a RDLE.
However, to fit accurately the spectrum, the constant of the vector interaction strength
must be reduced with respect to $\alpha_{g1}(0)$.
Moreover, the additional constant $\bar V_V$ to must be added to the vector potential.
Finally, a scalar or mass interaction is also strictly necessary to reproduce in detail
the charmonium spectrum.
Further investigation is necessary to establish a deeper connection between 
the effective bound state quark interaction and the phenomenology related 
to the QCD analysis.


\newpage
\vskip 0.5 truecm
\centerline{{\bf Acknowledgements}}
The author  gratefully thanks Prof. A. Deur and the other authors of Refs.
\cite{Deur05, Deur08, Deur22} for giving a complete numerical table
of the experimentally extracted $\alpha_{g1}(Q)$. The data of this table
are those shown in Fig. \ref{fig_1}.



\newpage
\begin{table*}

\caption{ 
Comparison between the experimental average values \cite{pdg22} 
 of the charmonium spectrum (last column)
and the theoretical results of the model.
All the masses are in MeV. 
The quantum numbers  $n$, $L$, $S$ and $J$, introduced in Ref. \cite{rednumb},
respectively
represent the principal quantum number, the orbital angular momentum, the spin 
and the total  angular momentum.
The results of the columns ``Scalar" and ``Mass" respectively refer to 
the scalar (S) and mass (M) interaction.
A line divides the resonances below and above the open Charm threshold.
At the bottom, the quantity $\Theta$, in $MeV$, defined in Eq. (\ref{qual}),
 gives an indication of the quality of the fit.  }

\begin{center}
\begin{tabular}{ccccc}
\hline
\hline \\
Name & $n^{2S+1}L_J$  & Scalar &  Mass & Experiment          \\
\hline \\
$\eta_c(1S)$    &  $1^1 S_0 $     & 2989   & 2994    & 2983.9   $\pm$  0.4   \\
$J/\psi(1S)$    &  $1^3 S_1 $     & 3100   & 3114   & 3096.9   $\pm$  0.006 \\
$\chi_{c0}(1P)$ &  $1^3 P_0 $     & 3418  & 3407   & 3414.71  $\pm$ 0.30   \\
$\chi_{c1}(1P)$ &  $1^3 P_1 $     & 3498  & 3494    & 3510.67  $\pm$ 0.05   \\
$ h_c(1P)$      &  $1^1 P_1 $     & 3511  & 3510    & 3525.38  $\pm$ 0.11   \\ 
$\chi_{c2}(1P)$ &  $1^3 P_2 $     & 3558   & 3564    & 3556.17  $\pm$ 0.07   \\
$\eta_c(2S)$    &  $2^1 S_0 $     & 3631   & 3626    & 3637.5   $\pm$ 1.1    \\
$\psi(2S)$      &  $2^3 S_1 $     & 3675  & 3677    & 3686.10 $\pm$ 0.06 \\
\\
\hline \\
$\psi(3770)$&     $1^3 D_1 $     &  3791 & 3784  & 3773.7   $\pm$ 0.4 \\  
$\psi_2(3823)$&   $1^3 D_2 $     &  3823 & 3819  & 3823.7   $\pm$ 0.5  \\
$\chi_{c1}(3872)$&$2^3 P_1 $     &  3898 & 3891  & 3871.65  $\pm$ 0.06 \\
$\chi_{c2}(3930)$&$2^3 P_2 $     &  3932 & 3933  & 3922.5   $\pm$ 1.0  \\
$\psi(4040)$&     $3^3 S_1 $     &  4017 & 4018  & 4039     $\pm$ 1     \\
%
$\chi{c1}(4140)$& $3^3 P_1 $     &  4153 & 4151  & 4146.5  $\pm$ 3.0    \\
$\psi(4230)    $& $4^3 S_1 $     &  4222 & 4227  & 4222.7   $\pm$ 2.6     \\
$\chi{c1}(4274)$& $4^3 P_1 $     &  4284 & 4292  & 4286    $\pm$ 9      \\
\\
\hline
\hline \\
$ \Theta   $          &$ ~$& $36.0 $&$38.0  $& \\  
\\
\hline\\
~\\
~\\
~\\
~\\

\end{tabular}
\end{center}
\label{tabspectr}
\end{table*}

\begin{table*} 
\caption{
Numerical values of the parameters of the model; for more details see
Sect. \ref{charmspectr}. The quark mass
$m_q $ is fixed at the value of Ref. \cite{pdg22}. 
The constant $\alpha_V(0)$ represents the strength of the vector interaction; 
$\bar V_V$ is the additional constant of the vector interaction.
The parameters of the scalar or mass interaction $\bar V_X$
and $r_X$ have the same values of Ref. \cite{scalint}.
         } 
\begin{center}
\begin{tabular}{lllll}
\hline 
\hline \\   
             &             &           &  Units     \\ 
\hline \\
 $m_q $      &  $1.27$     &            &   GeV   \\
\hline \\
%
%
             &   Scalar    & Mass       &             \\ 
\hline \\   
$\alpha_V(0) $&  $2.030  $ &$1.946 $ &    \\
%
$\bar V_V $  &  $1.837$    &$1.843$     &   GeV   \\
\hline\\
$\bar V_X $  &  $0.735$    &  $0.735$    &   GeV   \\
 $r_X      $&   $1.849$    &$1.846$     &  fm      \\     

\hline\\

\end{tabular}
\end{center}

\label{tabpar}
\end{table*}

\end{document}